%% file: ImpurityParallelDynamicsArticleV2.tex
\documentclass[10pt]{iopart}
\usepackage{iopams}
\usepackage{graphicx}
\usepackage{color}
\usepackage[utf8]{inputenc}
\graphicspath{{./Figures/}}

\newcommand{\bb}{\mathbf{b}}

\newcommand{\bB}{\mathbf{B}}
\newcommand{\bR}{\mathbf{R}}
\newcommand{\bh}{\boldsymbol{h}}

\newcommand{\bp}{\boldsymbol{\pi}}
\newcommand{\bu}{\mathbf{u}}
\newcommand{\bE}{\mathbf{E}}
\newcommand{\epol}{\mathbf{e}_\theta}
\newcommand{\etor}{\mathbf{e}_\phi}
\newcommand{\fsa}[1]{\left\langle{#1}\right\rangle}
\newcommand{\ps}{Pfirsch-Schlüter }
\newcommand{\eqref}[1]{(\ref{#1})}
\newcommand{\rev}[1]{#1}
\newlength{\colwidth}
\begin{document}

\title{Parallel impurity dynamics in the TJ-II stellarator}

\author{J~A~Alonso,  I~Calvo,  T~Estrada, J~M~Fontdecaba, K~J~McCarthy,  F~Medina, B~Ph~Van~Milligen, M~A~Ochando,  J~L~Velasco and the~TJ-II~Team}
%\address{Laboratorio Nacional de Fusi\'on, CIEMAT, Av. Complutense 40, 28040 Madrid}
%\ead{arturo.alonso@ciemat.es}
%\author{J.~L.~Velasco}
%\address{Laboratorio Nacional de Fusi\'on, CIEMAT, Av. Complutense 40, 28040 Madrid}
%\author{I.~Calvo}
%\address{Laboratorio Nacional de Fusi\'on, CIEMAT, Av. Complutense 40, 28040 Madrid}
%\author{B.~Ph.~Van~Milligen}
%\address{Laboratorio Nacional de Fusi\'on, CIEMAT, Av. Complutense 40, 28040 Madrid}
%\author{M.~A.~Ochando}
%\address{Laboratorio Nacional de Fusi\'on, CIEMAT, Av. Complutense 40, 28040 Madrid}
%\author{K.~J.~McCarthy}
%\address{Laboratorio Nacional de Fusi\'on, CIEMAT, Av. Complutense 40, 28040 Madrid}
%\author{T.~Estrada}
%\address{Laboratorio Nacional de Fusi\'on, CIEMAT, Av. Complutense 40, 28040 Madrid}
%\author{P.~Medina}
%\address{Laboratorio Nacional de Fusi\'on, CIEMAT, Av. Complutense 40, 28040 Madrid}
%\author{J.~M.~Fontdecaba}
%\address{Laboratorio Nacional de Fusi\'on, CIEMAT, Av. Complutense 40, 28040 Madrid}
%\author{the~TJ-II~Team}
\address{Laboratorio Nacional de Fusi\'on, CIEMAT, Av. Complutense 40, Madrid 28040, Spain}
\author{M Landreman}
\address{Institute for Research in Electronics and Applied Physics, University of Maryland, College Park,
Maryland 20742, USA}
\author{F~I Parra}
\address{Rudolf Peierls Centre for Theoretical Physics, 1 Keble Road, Oxford OX1 3NP, UK}
\author{J~M García-Regaña}
\address{EUROfusion PMU, Boltzmannstr. 2, 85748 Garching, Germany 5\\
Max-Planck-Institut für Plasmaphysik, Boltzmannstr. 2, 85748 Garching, Germany}
\author{J. Geiger and the W7-X Team}
\address{Max-Planck-Institut für Plasmaphysik, Wendelsteinstr. 1, 17489 Greifswald, Germany}

\begin{abstract}
We review in a tutorial fashion some of the causes of impurity density variations along field lines and radial impurity transport in the moment approach framework. An explicit and compact form of the parallel inertia force valid for arbitrary toroidal geometry and magnetic coordinates is derived and shown to be non-negligible for typical TJ-II plasma conditions. In the second part of the article, we apply the fluid model including main ion-impurity friction and inertia to observations of asymmetric emissivity patterns in neutral beam heated plasmas of the TJ-II stellarator. \rev{The model is able to explain qualitatively several features of the radiation asymmetry, both in stationary and transient conditions, based on the calculated in-surface variations of the impurity density.}

%We find that the phase of the modelled equilibrium impurity density asymmetry is in qualitative agreement with the shift observed in the emissivity. In pellet injection experiments, the emissivity pattern is observed to oscillate after every injection. We show that this is caused by an oscillation in the phase of the impurity density asymmetry following the sudden change caused by the pellet particle deposition. 
\end{abstract}

% Uncomment for PACS numbers
%\pacs{00.00, 20.00, 42.10}
%
% Uncomment for keywords
%\vspace{2pc}
%\noindent{\it Keywords}: XXXXXX, YYYYYYYY, ZZZZZZZZZ
%
% Uncomment for Submitted to journal title message
%\submitto{\JPA}
%
% Uncomment if a separate title page is required
%\maketitle
% 
% For two-column output uncomment the next line and choose [10pt] rather than [12pt] in the \documentclass declaration
%\ioptwocol
%

\section{Introduction}

The pollution caused by impurities is a potentially serious problem in magnetic confinement fusion plasmas. Constituent elements of the first wall and divertor materials are sputtered by chemical and physical processes and can be transported across the peripheral open field line region into the confinement region, where efficient fusion energy production can only be realised under stringent purity conditions. Once impurities cross the boundary of closed field lines, the negative radial electric field characteristic of this region (particularly in stellarators) causes an inward convection of the impurity species, which gets further ionised and convected as it finds itself in more interior, higher electron temperature locations.  The accumulation of impurities will proceed until the diffusive flux balances the convective one. In this condition, the impurity density peaking (i.e. the normalised inverse impurity density gradient) expected from the neoclassical balance is found to be considerably large for relevant conditions and impurity species of next generation stellarator devices like W7-X \cite{MollenJoP2012, MollenPoP2015}. In present devices the accumulation can limit the discharge duration \cite{BurhennNF2009}, but operational conditions have been found in which the plasma impurity content is kept down to affordable levels to maintain a stable plasma discharge \cite{McCormickPRL2002, NakamuraPPCF2014}. However, the incomplete understanding of some of these operational regimes make their extrapolation to future devices and reactor conditions uncertain. Generally the observations of low impurity content in plasmas with negative radial electric field is attributed either to conditions of reduced impurity transport in the scrape-off layer region, resulting in lower influx into the confinement volume  \cite{FengNF2006, NakamuraPPCF2014} or to anomalies in the radial transport of impurities in the confinement region \cite{McCormickPRL2002, YoshinumaNF2009}.

% Turbulence and Phi1 asymmetries as causes of anomalies
Recent theory and modeling  efforts have investigated both turbulent \cite{MikkelsenPoP2014} and extended neoclassical mechanisms \cite{GarciaPPCF2013} in an attempt to identify the physics underlying these anomalies. However, the observation of outward convection of impurities in the `impurity hole' regime of the Large Helical Device has not found an explanation in these efforts so far. The extension of neoclassical theory essentially consists on the proper accounting for electrostatic potential variations along flux surfaces (usually denoted in the literature as $\Phi_1$, \cite{MynickPoF1984, BeidlerISHW2005}) in the impurity orbits. In particular this includes the relaxation of the assumption of constant kinetic energy along particle trajectories (the so-called mono-energetic approximation) and the addition of the radial $\mathbf{v}_{E1}$  drift to the magnetic $\nabla B$ perpendicular gyro-center drift.

% Asymmetries or parallel impurity density variations
The experimental study of the variations of electrostatic potential along flux surfaces is difficult, as they are expected to be small in comparison with the radial variation of the electrostatic potential. In the plasma edge of the TJ-II stellarator, recent studies with Langmuir probes showed clear differences in the floating potentials of independent pairs of measurements on the same flux surface \cite{PedrosaNF2015}. These differences were seen to agree only qualitatively with kinetic simulations. Nevertheless, some observations of asymmetric impurity density could be the manifestation of those variations \cite{IngessonPPCF2000}, possibly augmented by heating-induced trapped particle populations. Besides parallel electric fields, inertia and friction with main ions are possible causes of medium and high $Z$ impurity density asymmetries (i.e. variations along flux surfaces). Reports on asymmetric radiation emissivity patterns are abundant both in the tokamak  and the stellarator scientific literature (see e.g. \cite{GiannonePPCF2003, KloseISHW2003, MarrPPCF2010, ReinkePPCF2012, ReinkePoP2013, ViezzerPPCF2013, ArevaloNF2014, AngioniPoP2015, CassonPPCF2015} to cite but a few).

In the first part of this article we briefly review these possible causes of impurity density variation in a tutorial fashion within a fluid framework. In the second part we apply the fluid modeling to recent TJ-II observations of off-centred emissivity patterns in tomographically inverted measurements of AXUV (Absolute eXtended UltraViolet) diode arrays. We find that, for the relatively cold and dense plasmas typically produced in neutral beam-heated TJ-II discharges, friction-driven impurity density asymmetries can qualitatively explain the observed shift in the radiation. The existence of these asymmetries and the ability of the fluid model to capture the essential physics is further corroborated by observations of transient oscillations of the emissivity pattern. These are reproducibly observed after pellet injections and can be understood as an oscillation in the phase of the impurity density asymmetry caused by the sudden breakdown of the parallel impurity force balance.

\section{Impurity parallel fluid dynamics and equilibrium density variations along flux surfaces}

To guide our discussion of possible causes and consequences of impurity density variations along flux surfaces we will use the particle and momentum conservation equations for an impurity species denoted by $z$, 
\begin{equation}
\frac{\partial n_z}{\partial t} + \nabla\cdot n_z \bu_z = S_z~,\label{eq:pce}
\end{equation}
\begin{equation}
\eqalign{
m_z \left(\frac{\partial n_z\bu_z}{\partial t} + \nabla\cdot n_z\bu_z \bu_z \right) + \nabla p_z + \nabla\cdot\bp_z \cr =Ze n_z\left(\bE + \bu_z\times\bB\right) + \bR_z~,}\label{eq:mbe}
\end{equation}
where all variables have their usual meaning (see e.g. \cite[Chapter 2]{CollisionalHelanderSigmar}). We note that the subindex $z$ denotes a single charge state of a certain impurity species. 
%For the main plasma ions (Hydrogen or Deuterium) and typical conditions found in magnetic confined fusion plasmas, the source term in the RHS of equation \eqref{eq:pce} can be neglected in the confinement volume. Within this volume,  electron density and temperatures are typically such that the neutral density and the recombination rate $H ^+\to H^0$ are small compared to the plasma density and the inverse transport time scale \rev{respectively}. However, in general the source term needs to be retained in the transport equations for the individual charge states of an impurity species, also within the confinement volume.
%
The radial transport equation is obtained from the flux surface average of equation \eqref{eq:pce} 
\begin{equation}\label{eq:trans}
\frac{\partial\fsa{n_z} }{\partial t} + \frac{1}{r}\frac{d}{dr} r \fsa{ n_z\bu_z\cdot\nabla r} = \fsa{S_z}~,
\end{equation}
where the radius $r$ is defined in terms of the flux surface volume as $r = (V/2\pi^2R_0)^{1/2}$, with $R_0$ the major radius. The term inside the radial derivative is proportional to the radial flux $\Gamma_z  = \fsa{n_z \bu_z\cdot\nabla r}$. 

For the case of a strongly magnetised plasma the fluid velocity perpendicular to the field line is given by the sum of the $E\times B$ and the diamagnetic flows, with corrections at least a factor of $\rho_{z}^*\ll 1$ (Larmor radius over system size) smaller. Therefore the fluid velocity can be written as
\begin{equation}\label{eq:uz}
\eqalign{
\bu_{z} = \left(\frac{d\Phi}{d\psi} + \frac{1}{Zen_z}\frac{d p_z}{d\psi}\right)\frac{\bB \times \nabla \psi}{B^2} & + \bu_{z\|} \cr &+ O(v_{tz}(\rho_{z}^*)^2 ),}
\end{equation}
where we have retained only the part of the electrostatic potential $\Phi$ and the pressure $p_z$  that are constant on flux surfaces, given by $\psi(\textbf{x}) = c$. Our flux coordinate $\psi$ is chosen to be the magnetic flux through the poloidal cross section of a flux surface divided by $2\pi$, $\Psi_T = 2\pi\psi$. The parallel velocity is given as $\bu_{z\|} = (\bu_z\cdot \bb) \bb$, with $\bb = \bB/B$ and the thermal velocity $v_{tz}$ is defined as $v_{tz} = \sqrt{T_z/m_z}$. We note that in particular the radial impurity particle flux in equation \eqref{eq:trans} is contained in the last term of equation \eqref{eq:uz}. We then conclude that the source term in equation \eqref{eq:pce} is of the same size as the radial flux $\fsa{S_z}/\fsa{n_z} \sim (v_{tz}/L)(\rho_{z}^*)^2 $. In consequence, the divergence of the impurity flux along the surface is larger than the other terms in this equation by a factor $\sim\rho_{z}^{*-1}$ and needs to cancel by itself \rev{in stationary conditions},
\begin{equation}\label{eq:pce2}
\nabla\cdot n_z\bu_z = 0~,
\end{equation}
\rev{with $\bu_z$ given by the first order terms in equation \eqref{eq:uz}.}
For given profiles of electrostatic potential $\Phi$ and pressure $p_z$, this is an equation for $\bu_{z\|}$. Indeed,  as far as the density is approximately a flux function $n_z = \fsa{n_z}(1 + O(\rho_{z}^*))$, equation \eqref{eq:pce2} reduces to the incompressibility condition $\nabla\cdot\bu_z = 0$ and the resulting parallel flow is the \ps flow\footnote{The \ps flow is generally defined to carry no net toroidal flow, $\fsa{u_{z\|}B} = 0$, which sets the integration constant for the magnetic differential equation.}.
However, in this article we will be discussing conditions in which the impurity density variation is `large', meaning that the corrections to the constant part of the density, even if ordered small in $\rho_{z}^*$, carry prefactors that can make them order unity in those conditions (see e. g.\cite{BraunPoP2010, LandremanPoP2011}). 

The parallel projection of equation \eqref{eq:mbe} can be simplified and cast in a form that allows to identify the main causes of impurity density variation of collisional impurities. The collisional coupling of main ions and impurity tend to equilibrate their temperatures, i.e. $T_z = T_i$, and therefore enables to disregard the impurity temperature variation with respect to the variation in its density. The anisotropic part of the pressure tensor $\bb\cdot\nabla\cdot\bp_{z}$ is dropped in favor of the friction force $R_{z\|} = \sum_{z'} R_{zz'\|}$, as frequent collisions tend to reduce the pressure anisotropy of the impurity species.  The main contribution to friction is considered to be the one due to main ions, $R_{z\|}\approx R_{zi\|}$, therefore assuming that other impurity species and charge states are present in low numbers compared to main ions. Using these approximations \rev{in the stationary version of} equation \eqref{eq:pce2} we get
\begin{equation}\label{eq:Bnablanz}
\eqalign{
\bB\cdot\frac{\nabla n_z}{n_z} =\cr \frac{B}{n_zT_z}R_{zi\|} -\frac{1}{v_{tz}^2}\bB\cdot\left(\bu_z\cdot\nabla\bu_z\right)  - \frac{Ze}{T_z}\bB\cdot\nabla\Phi ~,}
\end{equation}
from where several causes of impurity density asymmetries can be identified.
%These can derived directly from the zeroth and first velocity moments of the Focker-Planck equation. 
%The assumption of collisional impurity species allows to drop the anisotropic part of the pressure tensor against the friction force.  

\subsection{Parallel friction with main ions}
\begin{figure}
\includegraphics[width=.9\colwidth]{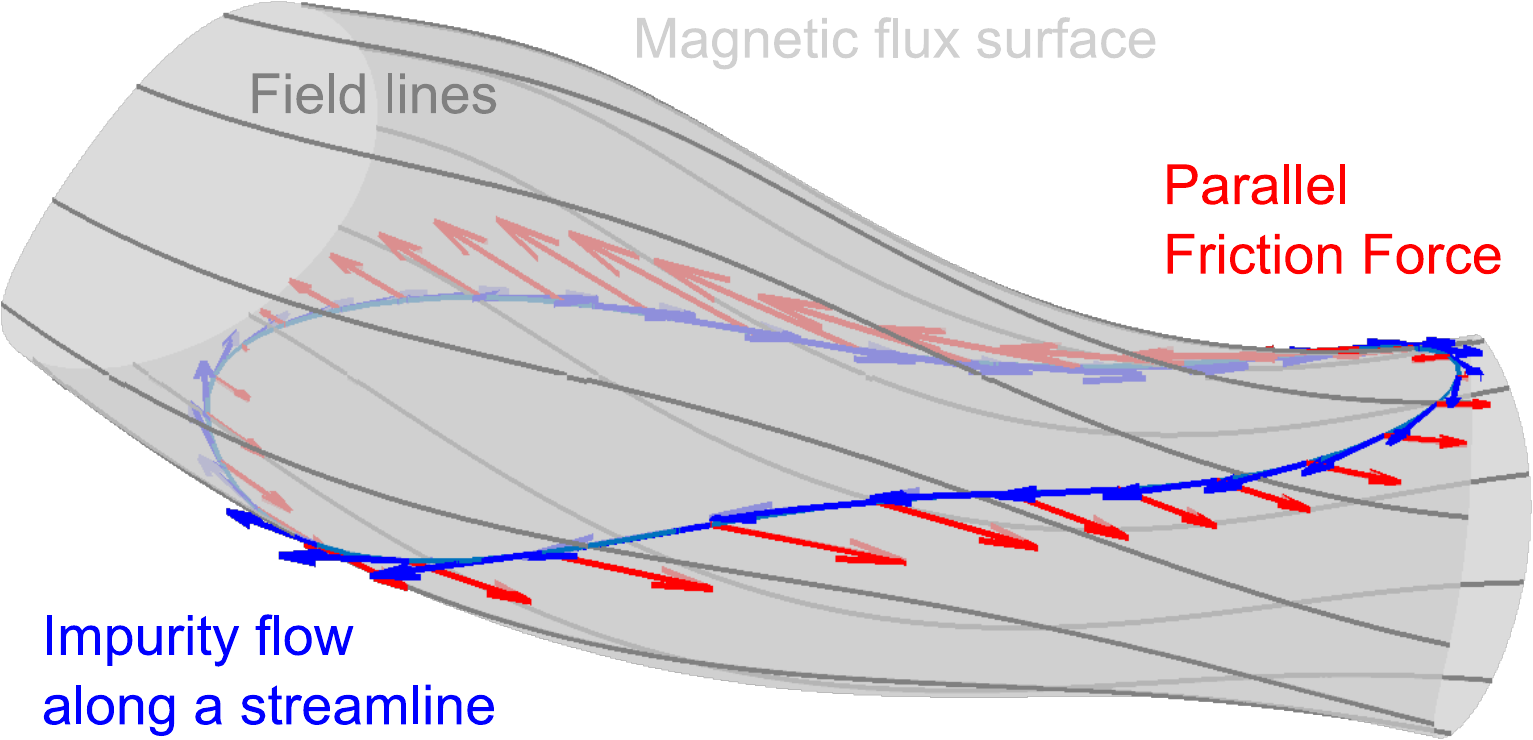}
\caption{Illustration of the friction force on impurities from the main ions along an incompressible streamline with zero net parallel flow. For the illustration we assume ion-root plasmas with $\omega_i = \Phi'(\psi) + p_i'(\psi)/en_i \approx 0$ and $\omega_z = \Phi'(\psi)$, so that main ions are assumed to move slowly along the streamlines.}
\label{fig:friction}
\end{figure}

Parallel friction with main ions is often approximated by the drag force on impurities due to the ion flow
\begin{equation}
R_{zi\|} \approx m_z n_z\nu_{zi}\left(u_{i\|} - u_{z\|}\right)~,
\end{equation}
which is appropriate for the qualitative comparison intended here. More exact expressions can be obtained with specific collision operators, which generally involve the ion distribution function (see e.g. appendix in \cite{ArevaloNF2014}). For the main ion flow we use an incompressible form (see \ref{sec:A1}). A streamline of that flow is shown in figure \ref{fig:friction}. If impurities distribute uniformly on a flux surface, their streamlines are coincident to those of the main ions, but the speed with which each species flows along the streamlines is different as determined by their different rotation frequencies ($\omega_s(\psi)$ in  \ref{sec:A1}). This causes friction between the species, whose parallel projection is largest where the parallel component of the streamline is largest. This occurs typically in the inboard and outboard sides of the flux surface as shown in figure \ref{fig:friction}. As a result impurities will tend to develop an up-down asymmetry as they tend to accumulate in (or be depleted from) the regions where the force cancels, $\nabla_\| \log n_z \propto R_{zi\|}$.

For the simulations presented here, the momentum transfer rate from the main ions to the impurity species was taken to be 
\begin{equation}\label{eq:nuzi}
\nu_{zi} = \frac{e^4\log \Lambda_c}{\pi\sqrt{\pi}\epsilon_0^2}\frac{Z^2 n_i}{v_{ti}^3m_i m_z}~,
\end{equation}
where a main ion charge $q_i=e$ and $m_z\gg m_i$ has been used (see e.g. \cite[Chapter 7]{Hazeltineframework}).
The first term on the RHS of equation \eqref{eq:Bnablanz} can then be written as $B\nu_{zi}^*(u_{i\|} - u_{z\|})/v_{tz}$, with the normalised friction coefficient $\nu_{zi}^*=\nu_{zi}/v_{tz}\sim Z^2/\sqrt{m_z}$ displaying a strong dependence on the charge $Z$ of the impurity species. Its dependence on the plasma parameters is $\nu_{zi}^*\sim n_i/T_i^2$.

The flux surface average of equation \eqref{eq:Bnablanz} yields $\fsa{(u_{i\|} - u_{z\|})B} = 0$ so that, in equilibrium, the net parallel flow of the impurity is that of the main ions. The cancellation of inertia upon flux surface averaging will be shown explicitly in section \ref{sec:inertia}.
% Something about the fact that we take upar average = 0
%

\subsection{Parallel inertia force}\label{sec:inertia}
\begin{figure}
\includegraphics[width=\colwidth]{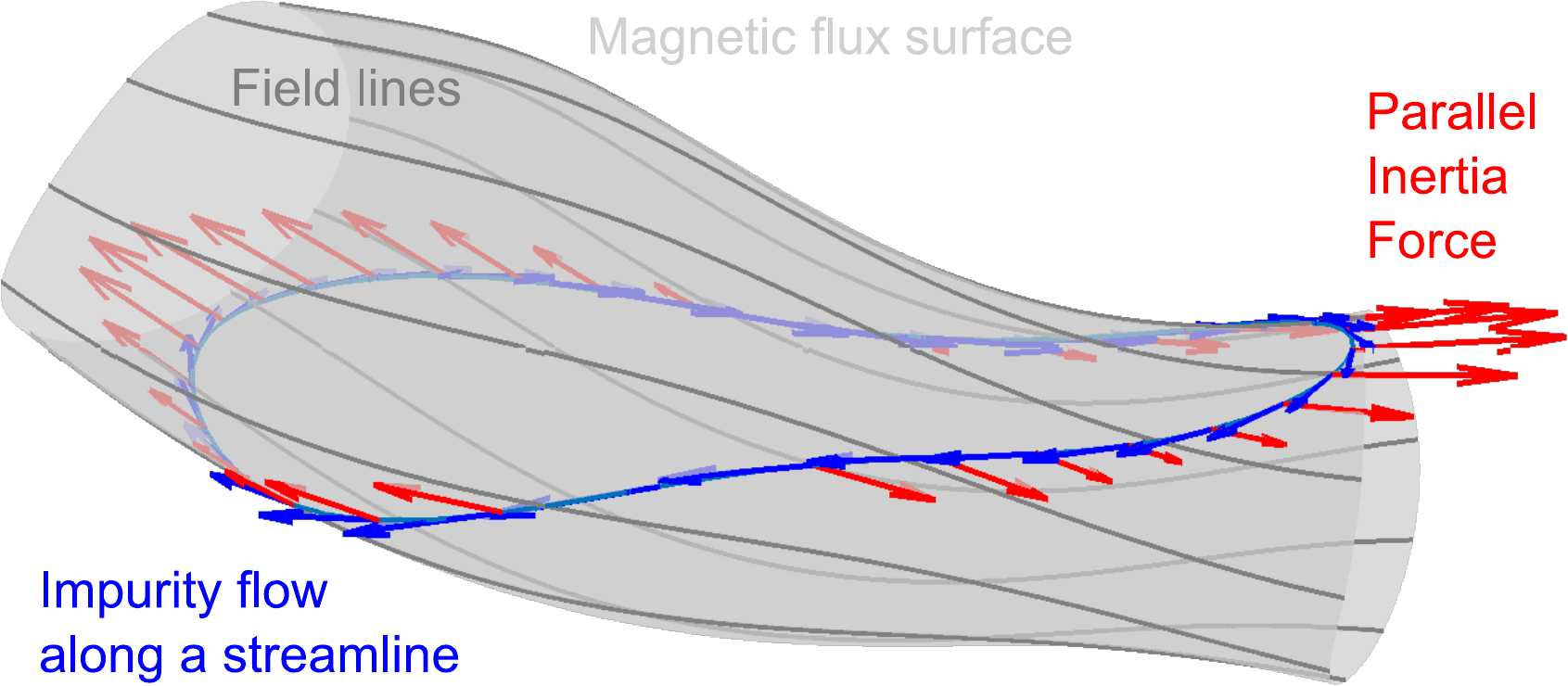}
\caption{Illustration of the inertia force on impurities. Parallel projection of the inertia (red arrows) along an incompressible flow line (blue) on a flux surface.}
\label{fig:inertia}
\end{figure}

The second term on the RHS of equation \eqref{eq:Bnablanz} is due to the inertia of the impurity flow, i.e. to the change of the velocity field along the streamlines of that same field. The parallel projection of this variation is a possible cause of impurity density asymmetries. To derive an expression of the parallel inertia that is easier to evaluate we write the impurity velocity field on a flux surface as
\begin{equation}\label{eq:umagz}
\bu_z = \omega_z(\psi)\etor + K_z(\psi, \theta, \phi)\bB~,
\end{equation}
where, $\etor$ is the toroidal basis vector of a general magnetic (straight-field line) coordinate system. Note that the first order flow in equation \eqref{eq:uz} can be written in this form (see \ref{sec:A1}).  Parallel inertia then reads 
\begin{equation}\label{eq:inertia}
\eqalign{
\bB\cdot(\bu_z\cdot\nabla)\bu_z = \cr \bB\cdot\nabla\left(\frac{K_z^2B^2}{2} -\frac{\omega_z^2g_{\phi\phi}}{2}\right)
+
\frac{\omega_z}{\sqrt{g}}\frac{\partial}{\partial\phi}\sqrt{g}u_{z\|}B~,}
\end{equation}
 in terms of the functions defined in equation \eqref{eq:umagz} and for an arbitrary magnetic coordinate system (see \ref{sec:A2} for details). The tokamak expression is readily recovered identifying $g_{\phi\phi}= R^2$ and canceling derivatives in the symmetry direction $\phi$.  From expression \eqref{eq:inertia} it is also evident that $\fsa{\bB\cdot(\bu_z\cdot\nabla)\bu_z} = 0$, i.e. parallel inertia does not exert a net force on impurities in a flux surface average sense. The parallel inertia force along a streamline is depicted in figure \ref{fig:inertia} for an incompressible streamline. 
Inertia is independent of the sign of velocity $\bu$, grows linearly with the mass of the impurity species and is strongest where the parallel \ps flow changes sign. \rev{It cancels at the outboard side with negative parallel derivative. Consequently, this force causes an initial redistribution of impurities from the inboard to the outboard side}\footnote{\rev{More precisely, the inertia force of an incompressible stream line is stellarator anti-symmetric and its parallel derivative at the minimum-$B$ center of symmetry (usually labeled $\theta =0, \phi=0$) is negative. The result of balancing this term with the parallel density gradient in \eqref{eq:Bnablanz} is then a stellarator symmetric impurity density $n_z$ with a maximum at $\theta=0, \phi=0$.}}. This is in fact the common wisdom about the effect of a strong toroidal rotation in a tokamak. Here we shown that this is also the case for the poloidally closed streamlines of the flow given by the sum of the perpendicular and parallel \ps flow. 

\begin{figure}
\includegraphics[width=\colwidth]{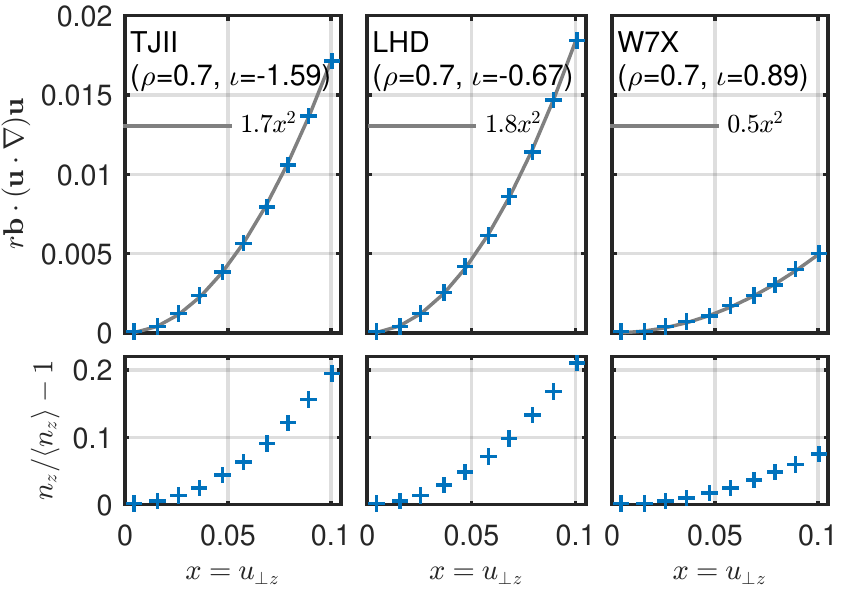}
\caption{Average values of inertia and equivalent density variations as a function of the normalised perpendicular velocity of the impurity species. The $\rho=0.7$ \rev{($\rho$ is the normalised radius defined as $\rho^2 = \psi/\psi_0$ with $\psi_0$ the flux coordinate of the last closed flux surface)} flux surfaces of three devices (TJ-II, LHD and W7X) are compared. The equivalent density variations are computed from the solution of $\nabla_\| \ln n_z = - \bb\cdot \left(\bu_z\cdot\nabla\right)\bu_z$. All velocities are normalised to thermal velocity.}
\label{fig:inertiascales}
\end{figure}

The dominant part of the inertia is due to the perpendicular advection of the parallel \ps flow, which has a characteristic $r^{-1}$ scale, i.e. 
\begin{equation*}
\bb\cdot \left(\bu\cdot\nabla\right)\bu \approx B \bu_\perp\cdot\nabla\left(\frac{u_\|}{B}\right)\sim \frac{\fsa{u_\perp}^2}{r}\frac{\partial}{\partial\theta}\left(\frac{u_\|}{\fsa{u_\perp}}\right)~.
\end{equation*}
The dimensionless derivative term is related to the \ps geometric factor in equation \eqref{eq:lambda}. It depends on the mode spectrum of $B$ and it is dominantly $m=1$ (this is visible from figure \ref{fig:inertia}). For illustration, in figure \ref{fig:inertiascales} we plot the inertia force and equivalent density variation of several magnetic configurations, as a function of the normalised perpendicular impurity velocity. In this figure it is noticeable that the reduction of \ps currents in the design of the W7-X magnetic geometry results also in a lower inertial force and smaller density variations compared to LHD or TJ-II.

To conclude we note that the net parallel rotation, i.e. the component of the parallel impurity flow $\bu_{z\|}$ of the form $V_{\|z}(\psi)\bB/\fsa{B}$, while possibly larger than the perpendicular and \ps components of the impurity flow, is much less efficient driving density variations.  The balance of pressure gradient and inertia gives, for this case,  $n_z/\fsa{n_z}-1 = ({V_{\|z}}/{v_{tz}})^2(\fsa{B^2}-B^2)/\fsa{B}^2$, which is smaller by roughly $\sim(r/R_0)^2$ than those shown in figure \ref{fig:inertiascales} for the same velocity. 

\subsection{Electrostatic potential variations}
The last term in the RHS of equation \eqref{eq:Bnablanz} is the force on impurity species due to the parallel electric field. It therefore involves the part of the electrostatic potential that is non-constant on flux surfaces. This part, often termed $\Phi_1$ in the literature, arises to preserve local charge neutrality and its computation requires to solve kinetic equations for ions and electrons. 
%: the adiabatic electron density response $n_e/\fsa{n_e}\sim e\Phi_1/T_e$, approximately balances the ion density variation from the first order (in an expansion in ion Larmor radius over system size) correction to the zeroth order constant ion density.  
We refer interested readers to the theory articles on the subject \cite{MynickPoF1984, HoPoF1987} as well as to the more recent work on kinetic simulations \cite{BeidlerISHW2005, GarciaPPCF2013} (cite Landreman et al. this issue). 
%More recently, the self consistent calculation of $\Phi_1$ has been included in the multispecies version of the \texttt{SFINCS} code [Cite Landreman, this conference]. We note that, while in general the approximation of an electron adiabatic density variation is a good one, specific low collisionality conditions can require to account also for the non-adiabatic part of electron distribution function in the quasineutrality equation.
% Rephrase this as ~ for those conditions in which the Neoclassical ion heat flux is larger the electron heat flux, the ion distribution function determines \Phi_1 

\subsection{Radial impurity transport}
\rev{The in-surface variations of impurity density or, more generally, of moments of the impurity distribution function are related to their radial transport.}
As is customary in the moment approach to neoclassical theory, flux-friction relations are obtained from the steady state  momentum balance equation \eqref{eq:mbe} dotted with a vector $\bh$  that can be defined by $\bh\times\bB = \nabla\psi$, $\nabla\cdot\bh = 0$ and $\fsa{h_\|B} = 0$. Flux surface averaging then yields
\begin{equation}\label{eq:flux}
\Gamma_z=  {\frac{1}{Ze}\fsa{\bh\cdot\bR_z}}
-
{\frac{1}{Ze}\fsa{\bh\cdot\nabla\cdot \bPi_z}}
-
{\fsa{n_{z}\bh\cdot\nabla\Phi}}~,
\end{equation}
where $\Gamma_z = \fsa{n_z\bu_z\cdot\nabla\psi}$ and  $\bPi_z = m_zn_z\bu_z\bu_z +{\bpi}_z$. The leading contribution to the pressure tensor has gyro-tropic form $\bpi_z = (p_{\|z} - p_{\perp z})\left(\bb\bb - \frac{1}{3}\mathbf{I}\right)$, so that $\fsa{\bh\cdot\nabla\cdot\bpi_z} = \fsa{(p_{\|z} - p_{\perp z})\bh\cdot\nabla \ln B}$. An expression analogous to \eqref{eq:flux}  can be obtained from the kinetic expression of the neoclassical flux and the drift-kinetic equation \cite{BraunPoP2010}, except for the classical diffusion --the perpendicular part of the first term in the RHS, $\fsa{h_\perp R_{\perp z}}$. We note that the neoclassical transport terms in equation \eqref{eq:flux} are essentially related to the spatial variations of  different moments of the impurity distribution function along flux surfaces. \rev{If, for the sake of illustration, the pressure anisotropy $(p_{\|z} - p_{\perp z})$ is constant on a flux surface, the second term in \eqref{eq:flux} vanishes upon flux surface averaging}.
The last term in \eqref{eq:flux} couples the variations of impurity density and electrostatic potential (related to the variations of main ion density). 
\rev{For a recent discussion on the possible effects of poloidal asymmetries on transport we refer the reader to \cite{AngioniNF2014b}.}

To finalize the heuristic discussion on some causes of impurity density variation, we note that the equilibrium impurity density variation and parallel flow need to fulfill both parallel force balance (equation \eqref{eq:Bnablanz}) and particle conservation (equation \eqref{eq:pce2}) so that the final phase and amplitude of the density variation will depend on the relative importance of the different forces and on the perpendicular velocity of the impurity. 
\rev{In the next section we solve simplified versions of these two coupled equations to model the observations of off-centered emissivity patterns in the TJ-II stellarator. The working hypothesis is that in-surface variations of emissivity are due to parallel variations of the density of the impurities that are the main cause of radiation. The model is derived in \ref{sec:A1}, \ref{sec:A2} and \ref{sec:model}. We note that electrostatic potential variations are not accounted for in this model.} This omission is briefly discussed in section \ref{sec:sum}.
We also note that in the previous discussions around figures \ref{fig:friction} and \ref{fig:inertia} we have disregarded the effect of a net parallel flow of main ions and impurities. The modeling results shown in the next section are relatively insensitive to the addition of a net parallel ion velocity.

%\section{Tomographic reconstruction of 2D emissivity patterns in TJ-II neutral beam heated plasmas}
\section{\rev{Modeling of radiation asymmetries in TJ-II neutral beam heated plasmas}}\label{sec:experiments}
\begin{figure}
\includegraphics[width=\colwidth]{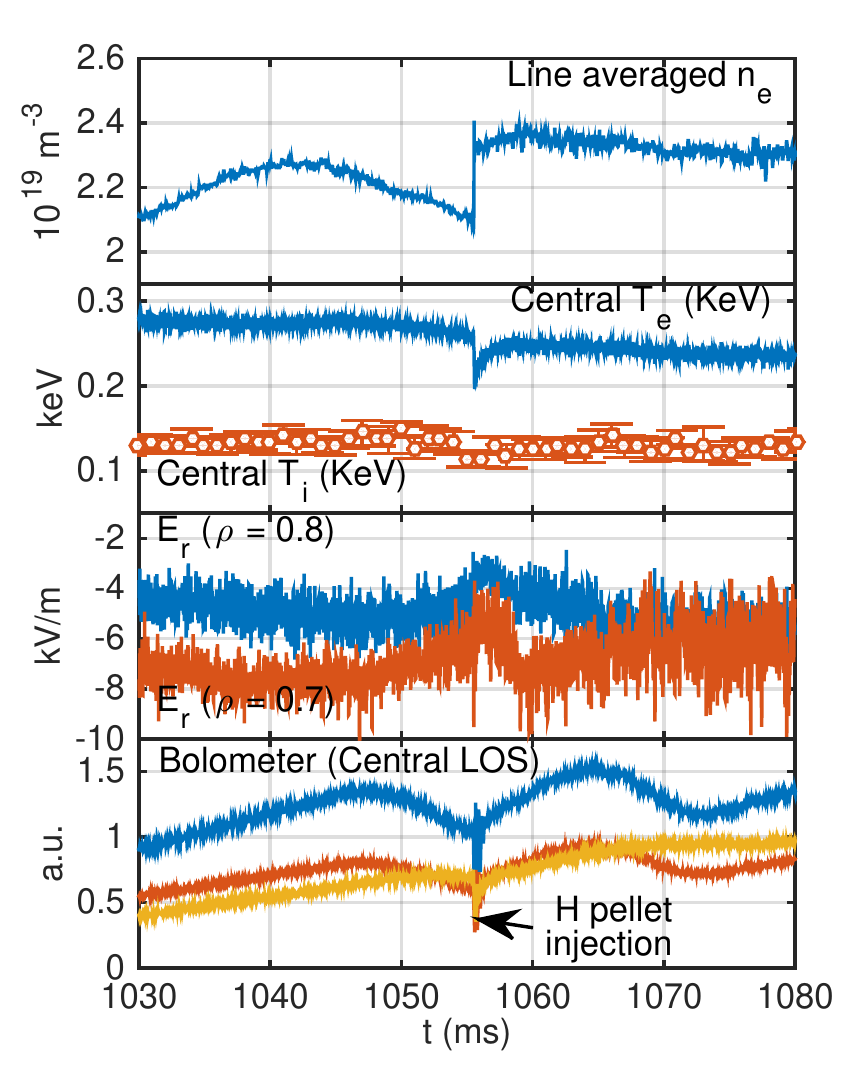}
\caption{Time traces of an NBI shot (TJII\#37486). From top to bottom: line averaged electron density from interferometer, central electron and ion temperatures from multifilter technique \cite{BaiaoRSI2010} and neutral particle analyser respectively, radial electric field at two radii from Doppler reflectometry and plasma radiation from AXUV-diode line integrated signals.}
\label{fig:timetraces}
\end{figure}

Typical neutral beam heated plasmas in TJ-II have densities in the range 1-4 $\times 10^{19} $m$^{-3} $, central ion and electron temperatures below $200$ eV and $400$ eV respectively. The radial electric field in these plasmas is seen to be negative across the minor radius. An example of the time traces of an NBI discharge is shown in figure \ref{fig:timetraces}. \rev{The plasma was heated with two NBI injections with nominal power of 600 kW and 400 kW with co- and counter-field injection directions respectively . Whereas no ion flow measurements were available in the analysed shots, the net ion parallel velocity for zero parallel momentum input is calculated to be $\sim 1.5$ km/s in the counter-field direction. Unbalanced NBI discharges with charge exchange measurements \cite{ArevaloNF2014} show values at mid radius below 10 km/s.} A hydrogen pellet injection is noticeable in the time traces around $t=1055$ ms. This triggers a transient response that will be the subject of section \ref{sec:pellet}. 

\begin{figure}
\includegraphics[width=\colwidth]{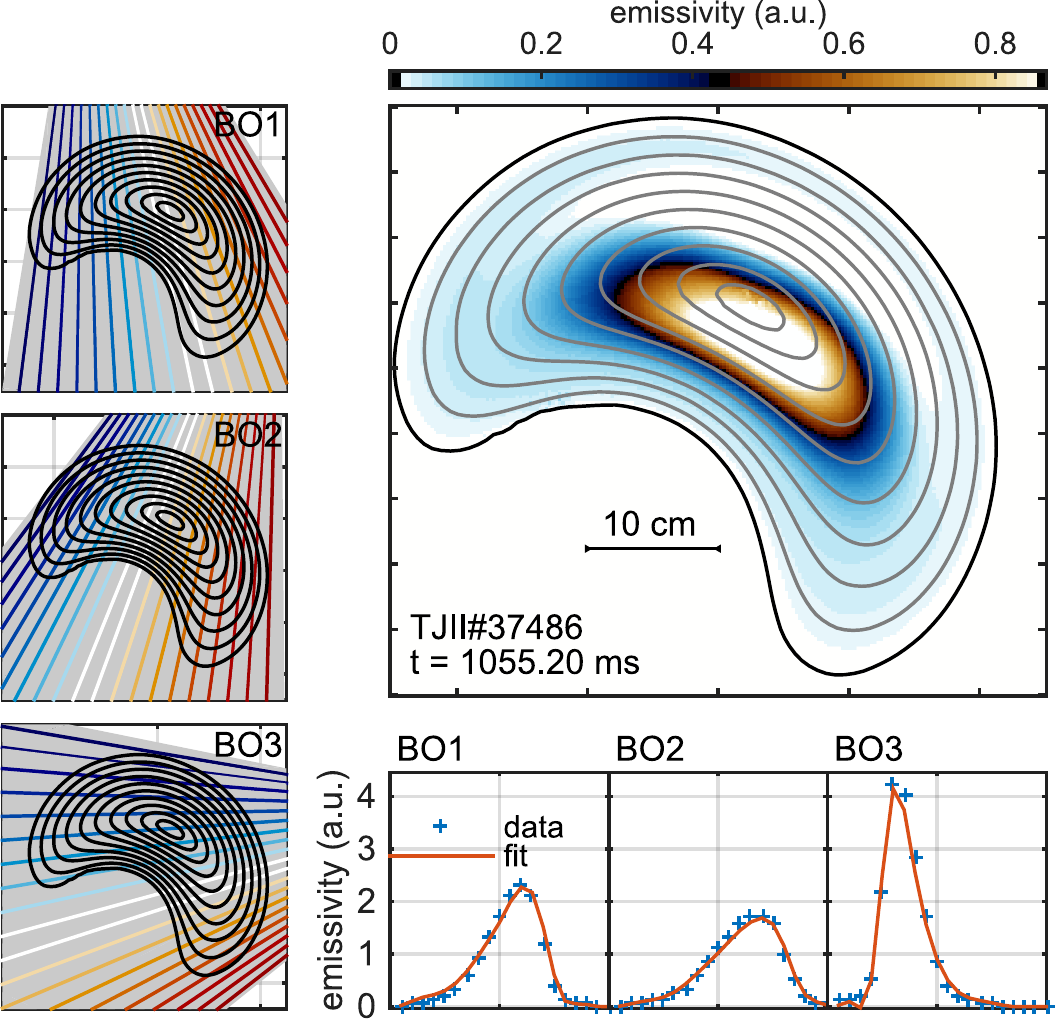}
\caption{Tomographic reconstruction of line-integrated emissivity measurements provided by three AXUV-diode arrays  (upper right axes) for the discharge shown in figure \ref{fig:timetraces}. The viewing lines of the arrays, named BO1 to 3,  are shown in the left column, and the comparison of the experimental data and the tomographic fit is shown in the lower three axes.}
\label{fig:tomo}
\end{figure}

\begin{figure}
\includegraphics[width=\colwidth]{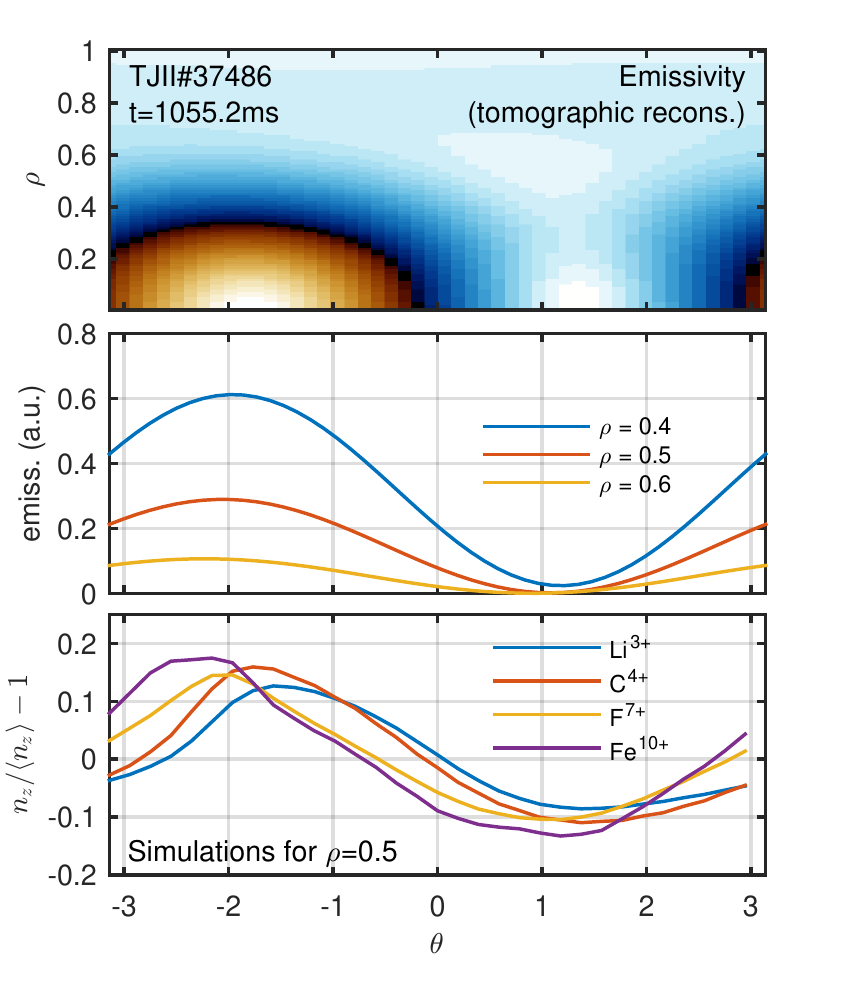}
\caption{Comparison between the tomographically reconstructed emissivity and the calculated equilibrium impurity density variations. The first plot shows the reconstruction from figure \ref{fig:tomo} in the $(\rho, \theta)$ plane. The angular dependence of the emissivity is shown in the middle plot for three radii. The bottom plot shows the variations of the density of typical TJ-II impurities calculated for $\rho=0.5$ at the cross-section of the measurements.}
\label{fig:CompSS}
\end{figure}

The tomographic reconstruction of the line integrated measurements provided by three AXUV-diode arrays before the pellet injection is shown in figures \ref{fig:tomo} and \ref{fig:CompSS}. The emissivity pattern shows a clear shift with respect to the magnetic surfaces. This off-centring is systematic in the database of 25 NBI discharges analysed, with line average densities spanning the range 2-4$\times 10^{19} $m$^{-3}$. Ion and electron temperatures do not change in the same proportion, as the increase in density also leads to a higher NBI absorption. \rev{For the discharge shown in figures \ref{fig:timetraces} and \ref{fig:tomo}, the inspection of the VUV spectrum shows dominant contributions from carbon and iron.} 

The tomographic inversion technique is based on parameter fitting with a Fourier/Bessel basis parametrisation of the emissivity in the poloidal/radial directions  (see \ref{sec:tomo}). For the cases shown here, five radial and two poloidal modes ($m=0,1$) were used in the reconstruction. The comparison of the experimental line integrated emissivity with the reconstructed one is shown in figure \ref{fig:tomo} (bottom) for the three cameras shown on the left hand side of the same figure. 
\rev{The relative error in the reconstruction (obtained as the root mean square of the fitting error of each cord divided by the root mean square of the cord signal values) is about 10\%, whereas symmetric ($m=0$) reconstructions cast  typical 30\% error. Increasing the number of modes further does not improve the reconstruction significantly.}

\rev{In order to model the observed stationary radiation asymmetry we look for a steady state impurity density variation which is solution of the model equations} \eqref{eq:dNdt} and  \eqref{eq:dKdt},  with physical parameters estimated from measurements (see table \ref{tab:params}). \rev{The radial electric field is calculated from neoclassical ambipolarity with experimentally constrained ion and electron profiles.} The comparison of the simulated density variations with the inverted emissivity variation is shown in figure \ref{fig:CompSS}.  The phase of the simulated asymmetry for several impurity species agrees well with the one observed in the emissivity. However, the amplitude of the simulated variations is below $\pm 20 \%$, whereas the emissivity shows very extreme differences between the maximum and the minimum values. 
\rev{Among the ingredients considered in the model,}
the main cause of impurity density asymmetries for these plasma conditions is friction with main ions which, in an average sense, is opposed by  the parallel inertia and pressure gradient forces. 
%The opposing effects of inertia and friction is the reason for the relatively small difference in the magnitude of the asymmetries between the different species in figure \ref{fig:CompSS} 

%For the modeling we choose an impurity species with $Z=4$ and $A = 12$ as representative of the impurities contributing to the radiation measured with the bolometer.  It is important to note that, in doing so, we do not imply that C$^{4+}$ is the species that contributes the most to the radiated power, but only that its parallel dynamics is representative.

%
\begin{table}
\begin{center}
\begin{tabular}{lcc}
\hline
Parameter & Value & Normalised\\
\hline
$v_{tz}$ & 28.1 km/s & 1\\
$R_0$ & 1.5 m & 1\\
$\tau_z = R_0/v_{tz}$ & 54 $\mu$s & 1\\
$v_{E}$ & 2.4 km/s & 8.5 $\times 10^{-2}$\\
$v_{i}^*$ & -1.2 km/s & -4.4 $\times 10^{-2}$\\
$\nu_{zi}$ & 12.4$\times 10^{4}$ s$^{-1}$ & 6.6 \\
\hline
\end{tabular}
\end{center}
\caption{Main ions (H$^+$) and impurity parameters used for the simulations. Values shown here are computed for C$^{6+}$ ($A=12, Z=6$). \rev{The values of the $E\times B$ and ion diamagnetic velocities are defined by $v_E = \frac{1}{\fsa{B}}\frac{d\Phi}{dr}$ and $v_i^* = \frac{1}{en_i\fsa{B}}\frac{d p_i}{dr}$. The flux surface averaged magnetic field is $\fsa{B} =0.95$T for the simulated surface $\rho = 0.5$. The main ion net parallel velocity, $\fsa{u_{i\|}B}$ and the impurity diamagnetic velocity $v_z^*$ are taken to be zero (note the $1/Z$ dependence of this velocity).}\label{tab:params}}
\end{table}
%
%
%\begin{figure}
%\includegraphics[width=\colwidth]{ForceBalance.pdf}
%\caption{Relative importance and direction of each term in the force balance equation. Bars in each group are calculated as the mean (over the poloidal and toroidal angles) of each of the three forces multiplied by the local sign of the reference force.}
%\label{fig:forcebalance}
%\end{figure}
%
\rev{An in-situ callibration of the diode viewing geometry and sensitivity has not been conducted. As a reference, the reconstruction of lower density ($\lesssim 1\times 10^{19}$m$^{-3}$) electron cyclotron heated discharges display lower levels of asymmetries with a different phase. The lower level of asymmetry is consistent with CXRS measurements of parallel  C$^{6+}$ flows \cite{ArevaloNF2013}. Therefore, whereas the effects of systematic spatial calibration and sensitivity errors cannot be fully determined at this point, they are unlikely to be the only cause of the reconstructed asymmetries seen in figure \ref{fig:tomo}. Additionally, the existence of angular impurity density variations will be show to explain the oscillatory behavior described in the next section.}

% Tomography
\subsection{Transient behavior of the asymmetry: phase oscillation after a pellet injection}
\label{sec:pellet}
\begin{figure}
\includegraphics[width=\colwidth]{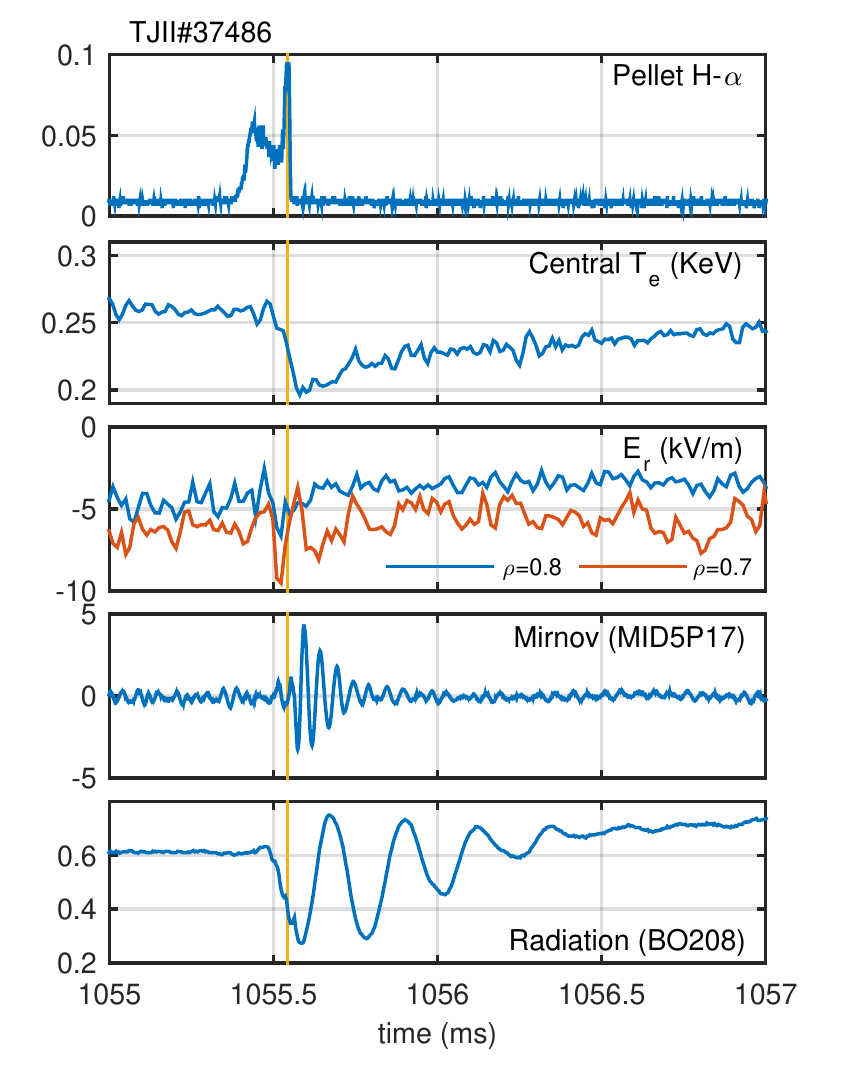}
\caption{Time traces of several plasma parameters around the injection of a hydrogen pellet. From top to bottom: H$_\alpha$ radiation from pellet ablation monitor, central electron temperature from multifilter technique, radial electric field from Doppler reflectometry, Mirnov coil signal and central AXUV-diode line. The excursion of the electric field towards negative values is also observed with HIBP diagnostic in more interior positions of similar discharges.}
\label{fig:closeup}
\end{figure}

The injection of cryogenic hydrogen pellets in the NBI plasmas discussed before is observed to cause an oscillation in the radiation monitors. \rev{The oscillation is observed reproducibly in the analysed database of 27 pellet injections with frequencies in a narrow window between 3 and 5 kHz.}. An example of such an injection is shown in figure \ref{fig:closeup}. The pellet ablation and the deposition of particles causes an increase in density (figure \ref{fig:timetraces}), a drop of the central electron temperature and transient decrease of the radial electric field towards more negative values. This is shown in figure \ref{fig:closeup} where the oscillation in the radiation monitors can also be seen. 
\rev{The pellet particle deposition occurs within $\rho=0.7$ and peaks about $\rho=0.5$. This leaves a flattened core density after the complete pellet ablation that evolves into a more peaked profile after 10 to 20 ms. This evolution has been found to be qualitatively consistent with neoclassical particle flux (cite J.L. Velasco et al., this issue).}

\begin{figure}
\includegraphics{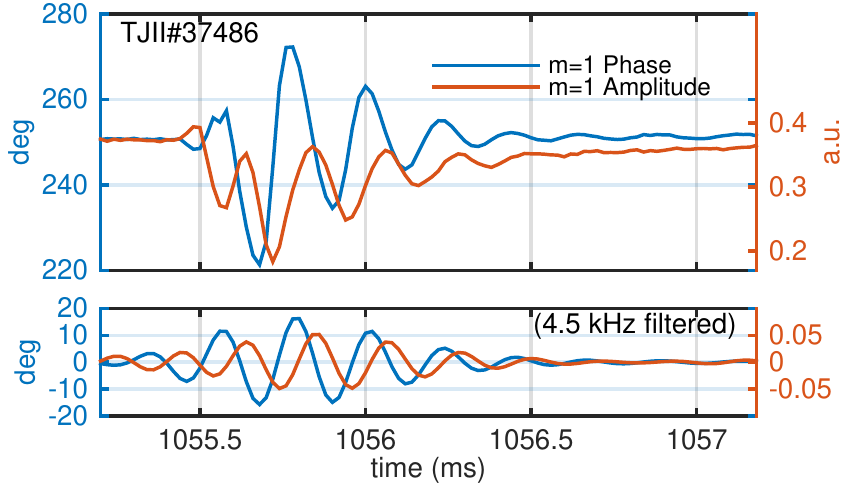}\caption{Time evolution of the $m=1$ phase and amplitude of the tomographically reconstructed emissivity after a pellet injection (same as in figures \ref{fig:closeup} and \ref{fig:comppellet}). The lower plot shows the $f\approx 4.5$ kHz components of the like-colored signals which show a $\pi/2$ phase difference between the phase and amplitude oscillations.\label{fig:phaseamplitude} We note that the lower frequency characteristic of the amplitude evolution also shows the effect of the electron density and  temperature evolution in the flux surface averaged emissivity (see figure \ref{fig:closeup}).}
\end{figure}

The temporal evolution of the tomographically reconstructed emissivity shows that the oscillation in the radiation monitors is mainly caused by an oscillation in the phase of the asymmetry with respect to the pre-pellet equilibrium situation shown in figure $\ref{fig:tomo}$. The time evolution of the phase and amplitude of the $m=1$ mode obtained from the tomographic inversion is shown in figure \ref{fig:phaseamplitude}. This behaviour has been simulated with the model equations \eqref{eq:dNdt} and \eqref{eq:dKdt}.
%

%tomographic inversion is shown in figure \ref{fig:comppellet}.
%In the same figure we show the simulated effect of the pellet injection on the impurity dynamics using the model equations \eqref{eq:dNdt} and \eqref{eq:dKdt}. 
%
\begin{figure}
\includegraphics[width=\colwidth]{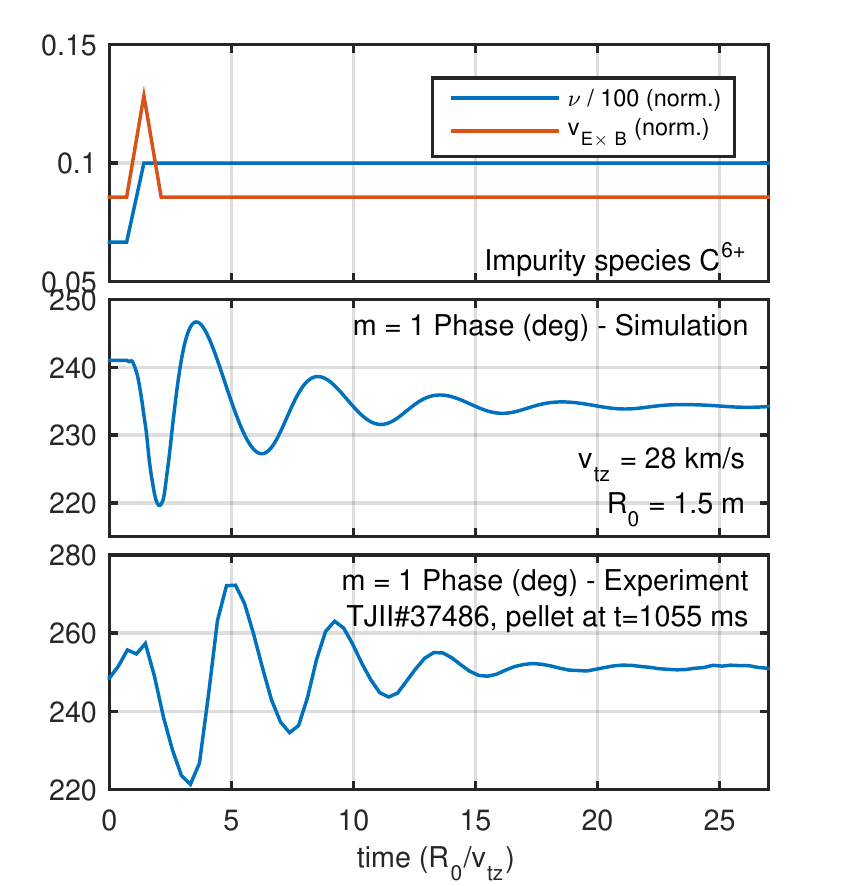}
\caption{Simulated pellet injection. Oscillation of the asymmetry phase and comparison with the observed phase oscillation.}
\label{fig:comppellet}
\end{figure}

The simulation starts from the C$^{6+}$ equilibrium density variation and parallel velocity fields obtained for the pre-pellet parameters. The normalised values of the friction coefficient and the $E\times B$ velocity are shown in the first time axes of figure \ref{fig:comppellet}. The effect of the pellet injection is synthesised as a sudden increase of the friction coefficient by 30$\%$, caused by the combination of the increase of density and the decrease of ion temperature (see equation \eqref{eq:nuzi}), and a transient excursion of the radial electric field towards more negative values as shown in figure \ref{fig:closeup}. The consistent evolution of parallel velocity field and density variation according to the model display a \rev{damped} oscillation in the phase of the $m=1$ component of the density variation similar to the one observed in the tomographic reconstructions. 
\rev{Both the frequency and the damping time are in close agreement with the observations. As discussed below, these are set by the $E\times B$ frequency and the inverse of the friction coefficient respectively. }
The observed amplitude of the oscillation is, nevertheless, somewhat larger than the simulated one. We note that the approximation of the angular dependence of the emissivity by that of the density of a single impurity species in the model is a crude one, \rev{and that considerable uncertainties exist in the ion temperature and other model parameters}.

%% comparison of simulations
The heuristic description of the oscillation on the asymmetry phase is the following. First a sudden change in the plasma parameters modifies the friction force and triggers a rapid adjustment of the parallel velocity field $K$ to restore force balance.  As a result, a part of the preexisting density variation $N$ is advected by the flow in the poloidal direction, as the precise compensation of the advection by the compression of the velocity field is lost. Both the rotating density variation and the static one are mainly $m=1$, and the superposition of the two causes an apparent oscillation of the phase of the $m=1$ component as observed in the experiments. 
\rev{This picture is also supported by the time evolution of the relative amplitude of the $m=1$ component obtained from the tomographic reconstructions (figure \ref{fig:phaseamplitude}), which shows a $\pi/2$ time phase with respect to the modulation in phase consistent with the behavior observed in the simulation: when the modulation of the angular phase of the asymmetry goes through zero, the rotating density perturbation is either aligned with the static one (amplitude shows a maximum) or is in phase opposition (amplitude shows a minimum).}

\begin{figure}
\includegraphics{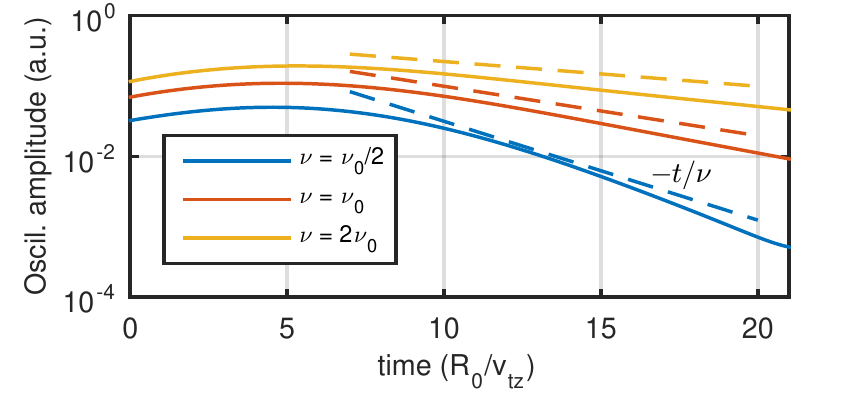}\caption{\label{fig:decay}Time envelope of the simulated oscillation for several values of the friction coefficient $\nu$. The condition $\nu=\nu_0$ corresponds to the simulations shown in  figures \ref{fig:comppellet} and \ref{fig:phaseamplitude}. Both the amplitude and the lifetime of the oscillation grow with increasing friction (see text).}
\end{figure}

\rev{The rotating density perturbation $\tilde{N}$ is mirrored by a perturbation in the parallel velocity field $B\tilde{K}\sim -\nu^{-1}\bb\cdot\nabla \tilde{N}$, so that parallel force balance is approximately fulfilled during the oscillation. This velocity response is responsible for the damping of the density perturbation through the second compressional term in the RHS of  the particle conservation equation \eqref{eq:dNdt}, with a characteristic normalised inverse time $\tau^{-1} \sim (k_\|R_0) ^2\nu^{-1}$, where $k_\|$ is the parallel wave-vector of the rotating density perturbation $\tilde{N}$. Figure \ref{fig:decay} shows the envelope of the simulated oscillations for three values of the friction coefficient $\nu$. The decay time agrees well with the above estimate with $k_\|R_0\sim 1$, $\tau\sim\nu$ (dimensionless units).
}

\section{Summary and discussion}\label{sec:sum}
We have shown the ability of a fluid modeling based on friction and inertia to reproduce some features of the observed asymmetric radiation emission in the TJ-II stellarator. 
The simulated phase of the equilibrium density variation of several impurities is in qualitative agreement with the tomographic reconstruction of the emissivity from AXUV-diode arrays. The magnitude of the simulated poloidal relative variation is however smaller that observed. We note that our model does not include the electrostatic potential variations, the computation of which requires the solution of the ion drift kinetic equation. However, its inclusion in the very similar model used in reference \cite{ArevaloNF2014}, yielded similar results for comparable plasma conditions. In that reference the asymmetry of the impurity density was assessed by comparing the inboard and outboard parallel velocities of C$^{6+}$. Whereas the model was not able to explain the sign of the measured flow variation, the predicted density asymmetry is qualitatively consistent with the 2D emissivity pattern shown here (see figures \ref{fig:tomo} and \ref{fig:CompSS} and figure 7 in reference \cite{ArevaloNF2014}).

In addition, we have studied the transient behaviour of the radiation asymmetry found after pellet injections. An observed oscillation in the phase of the radiation asymmetry was reproduced by the impurity fluid dynamic simulations and explained in terms of the advection of a \rev{decaying} $m=1$ asymmetry by the $E\times B$ flow. 
\rev{The oscillation frequency and damping time as well as the time lag between the oscillation phase and amplitude   are all consistent with the modeled evolution.}
According to this description, the oscillation is due to the presence of steady state impurity density asymmetries and the sudden breakdown of the advection/compression balance caused by the pellet injection. 
\rev{In this sense, the agreement found in the various characteristics of this transient behavior lends support to the existence of stationary impurity density asymmetries.}

\begin{figure}
\includegraphics{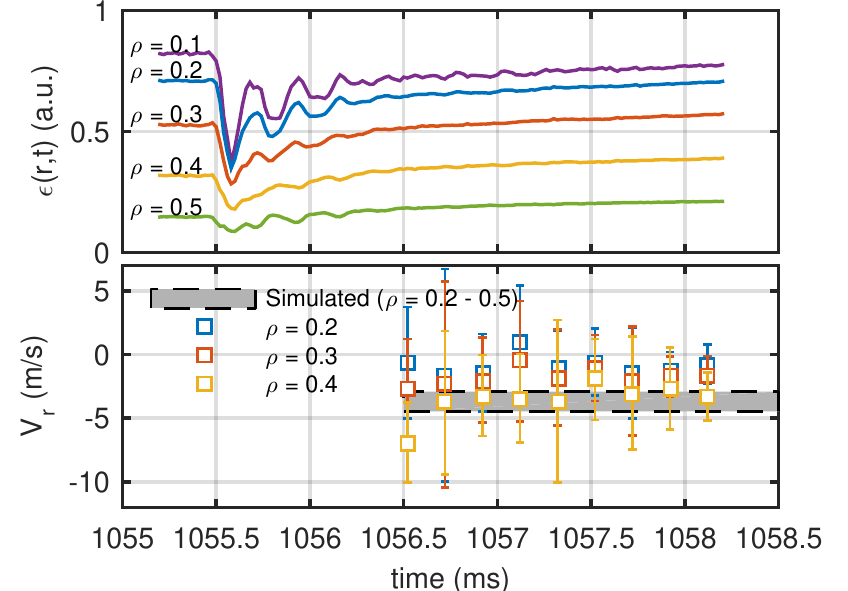}\caption{\label{fig:vrad}Time evolution of the emissivity during and after the pellet injection (top). Estimated characteristic radial velocity and comparison with the friction driven radial flux from the model (bottom).}
\end{figure}

\rev{As stated in the introduction, the radial impurity flux is related to the magnetic surface variation of different moments of the impurity distribution function, which makes the study of parallel impurity dynamics also relevant for its radial transport. The model used in this article allows to calculate the friction-driven radial impurity flux as given by first term in the RHS of equation \eqref{eq:flux} and friction of the form $\mathbf{R}_{z} = m_z n_z\nu_{zi}\left(\bu_{i} - \bu_{z}\right)~$. Whereas direct measurement of the radial flux of a specific impurity species are not available in TJ-II, the time evolution of the inverted emissivity profile $\epsilon(r,t)$ (poloidally integrated) can be used to obtain a rough estimate of the characteristic radial velocity of the impurities contributing to the radiation:
\begin{equation}
V_r(r) \sim \frac{-1}{r\epsilon(r,t)}\int_0^r \left( \frac{\partial \epsilon(r,t)}{\partial t} \right) r dr~.
\end{equation}
A few milliseconds after the pellet injection, central electron density is flattened by the particle source due to pellet ablation, while the electron temperature profile remains also flat as is characteristic of neutral beam heated plasmas in TJ-II (see Velasco et al., this issue). In these conditions, the radial dependence of the emissivity can be ascribed mainly to the change of the density of impurities that contribute to it. The radial velocity $V_r$ can be then considered a figure of the characteristic $\Gamma_I/\fsa{n_I}$ (radial flux over density) of those impurities. This estimate is shown in figure \ref{fig:vrad} for three central radial positions and compared to the modeled friction-driven flux  for C$^{6+}$. Both the calculated and estimated radial velocities yield values of a few negative (i.e. inwards) m/s. This figure shows $V_r$ for the specific pellet injection shown in figure \ref{fig:closeup}, but comparable values are obtained after other pellet injections in similar plasmas. Nevertheless,  due to the limitations of the modeling and experimental analysis here presented, this agreement should only be taken as indicative and needs to be complemented with dedicated flux measurements.
}

\rev{In this work we have studied experimental conditions with highly collisional impurities that are relevant for TJ-II NBI plasmas. Nevertheless, for stellarator-reactor relevant main ion collisionalities ($\nu_i\sim 10^{-3}$--$10^{-2}$), high-$Z$ species can approach the collisional regime ($\nu_z/\nu_i \approx Z^2\sqrt{m_i/m_z}$) at some radial locations.} 
In as much as the distribution of impurity density on a flux surface is related to the radial transport across it, the understanding of the different causes of impurity density and electrostatic potential variations are important to assess the instrumentality of those variations for controlling impurity accumulation in magnetically confined plasmas.

\section*{Acknowledgements}
We acknowledge useful discussions with P. Helander and C. Beidler. We thank C. Beidler also for reading the manuscript and providing useful comments.  We thank S. Satake and the NIFS Team for providing the VMEC equilibrium for the LHD. This work has been carried out within the framework of the EUROfusion Consortium and has received funding from the Euratom research and training programme 2014-2018 under grant agreement No 633053. The views and opinions expressed herein do not necessarily reflect those of the European Commission. This research was supported
in part by grants ENE2012-30832 and ENE2015-70142-P from Ministerio de Econom\'ia y Competitivad, Spain.
 
\appendix
\section{Expression of the flows in magnetic coordinates}\label{sec:A1}
In magnetic coordinates $(\psi, \theta, \phi)$,  with $\psi = \Psi_T/2\pi$, the magnetic field is expressed 
\begin{equation}\label{eq:B}
\bB = \nabla\psi\times\nabla(\theta - \iota\phi) = \frac{1}{\sqrt{g}}\left(\iota\epol + \etor\right)~,
\end{equation}
%
%$\epol := \frac{\partial \mathbf{r}}{\partial\theta} = \sqrt{g}\nabla\phi\times\nabla\psi$
where $\sqrt{g} = \left|\nabla\psi\cdot\nabla\theta\times\nabla\phi\right|^{-1}$ is the Jacobian of the coordinate system and $\iota = d\Psi_P/d\Psi_T$ the rotational transform. In terms of these definitions, the flow defined in equation \eqref{eq:uz} can be written as
\begin{equation*}
\bu_s = \hat\omega_s(\psi)\epol + \hat K_s(\psi, \theta, \phi)\bB~,
\end{equation*}
where $\hat\omega_s(\psi) = d\Phi/d\psi + (1/Z_sen_s)(d p_s/d\psi)$. This can be checked using  equation \eqref{eq:B} and the relation $\epol\times\bB = \nabla\Psi_T/2\pi = \nabla \psi$. Alternatively, one can choose to write it as
\begin{equation}\label{eq:umag}
\bu_s = \omega_s(\psi)\etor + K_s(\psi, \theta, \phi)\bB~,
\end{equation}
where now $\omega_s(\psi) = -\hat\omega_s(\psi)/\iota(\psi)$. We will work with form \eqref{eq:umag}, which is reminiscent of the tokamak split of flows in toroidal and parallel directions, with $K$ given in terms of the poloidal velocity $u^\theta = KB^\theta$. The expressions used in this article can therefore be directly adapted to the axisymmetric case identifying $\etor = R\vec{\boldsymbol{\varphi}}$, where $\vec{\boldsymbol{\varphi}}$ is the unit vector in the direction of the cylindrical coordinate line.

For the main ion flow we assume that density variations are negligible and impose $\nabla\cdot\bu_i = 0$ to get
\begin{equation}
K_i  = \omega_i(\psi)\left(\frac{\lambda}{B} - \frac{B_\phi}{\fsa{B^2}}\right) + \frac{\fsa{u_{i\|}B}}{\fsa{B^2}}~,
\end{equation}
where the function $\lambda/B$ is given by
\begin{equation}\label{eq:lambda}
\bB\cdot\nabla\frac{\lambda}{B} = -\frac{1}{\sqrt{g}}\frac{\partial\sqrt{g}}{\partial\phi}
\end{equation}
with $\fsa{\lambda B} = 0$.
For solving this magnetic differential equation and  evaluating other terms, it is convenient to use Boozer coordinates. In those coordinates $B_\phi = I(\psi)$ that is proportional to the poloidal current through a disk limited by a toroidal circuit over the flux surface. For the tokamak case, $\lambda/B = 0, B_\phi = RB_T$ and $K_i$ is a flux function. The form of the function $\lambda$ determines the streamlines of the incompressible flow which are also the current circuits of the diamagnetic $+$ \ps currents.

\section{Derivation of the parallel inertia (equation \eqref{eq:inertia})}\label{sec:A2}
We start from the common split
\begin{equation*}
\bB\cdot\bu\cdot\nabla\bu = \bB\cdot\nabla\frac{u^2}{2} + \bB\cdot(\nabla\times\bu)\times\bu~,
\end{equation*}
and work out the second term using the form \eqref{eq:umag}, 
%\begin{equation}
\begin{equation}\label{eq:t1}
\eqalign{\bB\cdot & (\nabla\times\bu)  \times\bu = (\nabla\times\bu)\cdot(\bu\times\bB)
\cr &= (\nabla\times\bu)\cdot(\omega\etor\times\bB)
\cr &= \nabla\cdot\left(\omega\bu\times(\etor\times\bB)\right)
\cr &= \nabla\cdot\left(\omega(\bu\cdot\bB)\etor-\omega(\bu\cdot\etor)\bB\right)
\cr &= \frac{\omega}{\sqrt{g}}\frac{\partial}{\partial\phi}\sqrt{g}u_\|B - \bB\cdot\nabla(\omega^2 g_{\phi\phi}+ \omega KB_\phi)~,}
\end{equation}
where $g_{\phi\phi} =\etor\cdot\etor$. 
The first term is simply
\begin{equation}\label{eq:t2}
\eqalign{\bB\cdot\nabla\frac{u^2}{2} =\cr \bB\cdot\nabla\left( \frac{\omega^2g_{\phi\phi}}{2} + \frac{K^2B^2}{2} + \omega K B_\phi \right)~,}
\end{equation}
and summing \eqref{eq:t1} and \eqref{eq:t2} we get
\begin{equation*}
\eqalign{
\bB\cdot(\bu\cdot\nabla)\bu =\cr \bB\cdot\nabla\left(\frac{K^2B^2}{2} -\frac{\omega^2g_{\phi\phi}}{2}\right)
+
\frac{\omega}{\sqrt{g}}\frac{\partial}{\partial\phi}\sqrt{g}u_\|B~.}
\end{equation*}
Note that this expression is valid for any straight-field line coordinates.

\section{Model equations for impurity density variations and parallel flow}\label{sec:model}
For the modelling of the observations presented in section \ref{sec:experiments} we use the impurity particle conservation equation and momentum balance including parallel pressure gradient, inertia  and friction. These provide coupled evolution equations for the two functions of the angular variables  $N=\log (n_z/\fsa{n_z})$ and $K = K_z$ as defined in equation \eqref{eq:umag}. 
\rev{Other parameter definitions can also be found in \ref{sec:A1}. The time-dependent version of the first order continuity equation \eqref{eq:pce2} reads in terms of these variables}
\begin{equation}\label{eq:dNdt}
\frac{\partial N}{\partial t} = -\omega\frac{1}{\sqrt{g}}\frac{\partial\sqrt{g}}{\partial \phi} - \bB\cdot\nabla K - \omega\frac{\partial N}{\partial \phi}- K\bB\cdot\nabla N ~,
\end{equation}
 % 
% %
%\begin{equation}
%\frac{\partial N}{\partial t} = -\nabla\cdot\bu -\bu\cdot\nabla N~.
%\end{equation}
 % 
%
\rev{As parallel force balance equation we use \eqref{eq:Bnablanz}, in the approximation $\bB\cdot\nabla\Phi = 0$, and include the time derivative of the parallel velocity to get}
\begin{equation}\label{eq:dKdt}
\eqalign{B^2\frac{\partial K}{\partial t} = &-B_\phi\frac{\partial \omega}{\partial t} - \bB\cdot\nabla N \cr
& -\bB\cdot\nabla\left(\frac{K^2B^2}{2} -\frac{\omega^2g_{\phi\phi}}{2}\right)\cr
& -\frac{\omega}{\sqrt{g}}\frac{\partial}{\partial\phi}\sqrt{g}(\omega B_\phi + KB^2)\cr
&+ \nu\left( (\omega_i-\omega)B_\phi + (K_i - K)B^2\right)~.}
\end{equation}
 In these equations, velocities are normalised to the impurity thermal velocity $v_{tz} = \sqrt{T_z/m_z}$, spatial scales are normalised to the characteristic system size $R_0$ and times are normalised to the transit time $R_0/v_{tz}$. 
The input parameters are  the ion and impurity rotation frequencies $\omega_i, \omega_z\equiv\omega$, the friction coefficient $\nu=R_0\nu_{zi}^*$ (see equation \eqref{eq:nuzi}) and the magnetic geometry. Note that equations \eqref{eq:dNdt} and \eqref{eq:dKdt} depend only on the relative density variation $n_z/\fsa{n_z}$ and not on the absolute value of the impurity density. The metric coefficient $g_{\phi\phi} = \etor\cdot\etor$. The axisymmetric version of these equations is readily obtained by identifying $B_\phi = RB_T$, $g_{\phi\phi} = R^2$ and removing all derivatives in the direction of symmetry $\partial_\phi f = 0$.

\section{Tomographic reconstruction technique}\label{sec:tomo}

A tomography program has been developed for the reconstruction of the local emission of radiation from a set of line-integrated measurements in a single poloidal plane ($\phi =$ constant). 
The local emission $E$ is expressed as a function of the VMEC \cite{vmec} flux coordinates 
$(\rho,\theta)$ in the form of a series expansion:
\begin{equation}
E(\rho,\theta) = \sum_{n,m}{C_{nm}f_{n}(\rho) \exp(im\theta)}
\end{equation}
%Separating out the $m = 0$ terms:
%\begin{equation}
%E(\rho,\theta) = \sum_{n=0}^N{C^0_{n0} f^0_{n}(\rho)} + \sum_{n,m=1}^{N,M}{f^1_{n}(\rho)\left [C^1_{nm}\cos(m\theta) + C^2_{nm}\sin(m\theta) \right ]}
%\end{equation}
For an optimal reconstruction process, one would like the expansion functions to form an orthogonal set (thus avoiding statistical collinearities). The orthogonality should apply with respect to the area integral, since this is what contributes to the measured emission. This leads to:
\begin{equation}
\int_0^1{f_{n}(\rho)f_{n'}(\rho) \rho d\rho} = c_n \delta_{nn'}
\end{equation}
where $\delta_{ij}$ is the Kronecker delta. The Fourier-Bessel functions satisfy this requirement 
\cite{vanMilligenRSI2011}.

The local emission $E(\rho,\theta)$ is calculated on a sufficiently fine grid $(\rho_i,\theta_j)$, $i = 1,\dots,N_\rho, j = 1,\dots,N_\theta$. 
The emission of a grid element (with area $da_{ij}$) to detector number $\nu$ is evaluated taking into account the distance $d^\nu_{ij}$ of the grid element $(i,j)$ to the detector:
\begin{equation}
I^\nu_{ij} = \frac{E(\rho_i,\theta_j)}{(d^\nu_{ij})^2} \delta^\nu_{ij} da_{ij}
\end{equation}
Of course, only grid elements lying inside the detector beam viewing angle are contemplated;
this is expressed by the factor $\delta^\nu_{ij}$ which equals 1 when the grid element lies inside the beam, and 0 otherwise (partial overlap between the viewing beam and the grid element is not contemplated since the grid is assumed to be sufficiently fine).
By integrating over all grid elements $(i,j)$, the total emission received by detector $\nu$, $I^\nu$ can be calculated.

Since the local emission $E(\rho,\theta)$ is expanded as a series, this integration can be performed for each mode separately. Labeling the modes $(n,m)$ by a single index $\mu$, we can write the relation between mode amplitudes and detected intensities as:
\begin{equation}
I^\nu = A^\nu_\mu C_\mu
\end{equation}
where the $C_\mu$ are the mode coefficients. This is a linear set of equations that can be solved using standard techniques, allowing fast recovery of the mode coefficients from the detector signals. 
The detector signals $I^\nu$ are assigned weights $w_\nu$ that are inversely proportional to the square of the measurement error of measurement $I^\nu$. 
Thus, any signal $\nu$ can be excluded from contributing to the result by setting $w_\nu = 0$.

The regression technique used is a standard combined Gauss-Newton and modified Newton gradient descent algorithm that minimizes both the reconstruction error and the mode amplitudes.

%\bibliographystyle{iopart-num}
%\bibliography{C:/Users/aalso/RZGshare/References/MyNewBib}{}
\input{ImpurityParallelDynamicsArticleV2.bbl}

\end{document}

%% file: ImpurityParallelDynamicsArticleV2.bbl
 \newcommand{\noop}[1]{}
\providecommand{\newblock}{}